# Observation of large intrinsic anomalous Hall conductivity in polycrystalline Mn3Sn films

Wafa Afzal*[1,3], Zengji Yue[1,3], Zhi Li[1,3], Michael Fuhrer[2,3], Xiaolin Wang*[1,3]

1. Institute of Superconducting and Electronic Materials (ISEM), Australian Institute for Innovative Materials (AIIM), University of Wollongong, Squires Way, North Wollongong, 2500, Australia.

2. School of Physics and Astronomy, Monash University, Clayton, Victoria, Australia

3. Future Low Energy Electronics Technologies (FLEET), Australian Research Centre (ARC), Australia.

**Abstract**

We report the observation of anomalous Hall effect in $Mn_3Sn$ polycrystalline thin films deposited on Pt coated $Al_2O_3$ substrate with a large anomalous Hall conductivity of $65(\Omega cm)^{-1}$ at 3K. The Hall and magnetic measurements show a very small hysteresis owing to a weak ferromagnetic moment in this material. The longitudinal resistivity decreases sufficiently for the thin films as compared to the polycrystalline bulk sample used as the target for the film deposition. The anomalous Hall resistivity and conductivity decreases almost linearly with the increase in the temperature. A negative magnetoresistance is observed for all the measured temperatures with the negative decrease in the magnitude with the increase in temperature.

**Introduction**

The Anomalous Hall Effect (AHE) was only understood for the ferromagnetic materials for a very long time. Being proportional to the net magnetization, it was thought that it vanishes for the antiferromagnets where there is no net magnetization. The recent studies, however, dictate the presence of an anomalous hall effect even in the absence of net magnetization for certain symmetries [1]. Anomalous Hall Effect has been observed for non-collinear antiferromagnets [2-4] and now it is understood as an intrinsic quantum mechanical phenomena dependent on the berry curvature of the bands in the momentum space that represent topologically non trivial phases [5, 6]. The Berry curvature takes on a non-zero value in the time symmetry breaking systems [7], so it does not appear in antiferromagnetic system as the net magnetization vanishes. However, for non-collinear antiferromagnets, the Berry curvature doesn't vanish and may give rise to a high AHE. This effect was not well understood until recent observation of a large anomalous hall effect for non collinear AFMs even at the room temperature [8, 9].

$Mn_3Sn$ was theoretically described as a material hosting Weyl nodes and an anomalous Hall conductivity [10, 11]. Large AHE has been reported for the single crystals and the thin films of $Mn_3Sn$ [4, 12-15]. The presence of Weyl fermions has been detected through a combination of ARPES studies along with the first principle calculations. The experimental evidence for the presence of Weyl nodes was presented with the help of angle resolved spectroscopy by directly observing the Weyl nodes using single crystals [16]. $Mn_3Sn$ is one of the very first antiferromagnetic Weyl semimetals and the reported transport studies also provided significant evidence for the Weyl fermions in this material. Positive magneto-conductance for the parallel direction of magnetic field with the current and the negative for the perpendicular shows positive signatures of the chiral anomaly in the material [17]. Magneto-optical Kerr effect (MOKE) has also been reported for this material also being the first MOKE report on antiferromagnets. A large zero field Kerr rotation angle at room temperature, as large as found in ferromagnets, was reported for $Mn_3Sn$ single crystals[18, 19]. This attracted significant interest in this material and other Mn based non-collinear antiferromagnets ($Mn_3X$ series, X=Ge, Sn, Pt, Ir) exhibiting strong anomalous Hall Effect. A fictitious magnetic (Berry) field of more than 200T is generated by the band structure of these materials in order to explain the exhibited AHE [3].

$Mn_3Sn$ has a hexagonal crystal structure with lattice parameters a=b=5.592A, c=4.503A and belongs to the space group $P6_3/mmc$ (194). It is an antiferromagnet that exhibits non collinear ordering of Mn moments below the Neel temperature of 420K.[20] An antiferromagnetic moment of 3 $\mu_B$ and a very small ferromagnetic moment of 0.002 $\mu_B$ appears per Mn atom. The basal plane (0001) of $Mn_3Sn$ consists of Mn atoms arranged in a kagome lattice with the non collinear inverse triangular spin structure with the neighbouring moments aligned at 120° and the Sn atoms lie at the centre of the hexagons formed by Mn atoms. The small ferromagnetic moment arises due to the slight canting of the triangular Mn spins towards their easy axis and the non-stoichiometry of the compound where Mn atoms take some of the Sn positions. The compound is reported to be stable in a slight excess of the Mn atoms[8].

This work presents a comprehensive study of the polycrystalline thin films of $Mn_3Sn$. We synthesized a polycrystal to be used as the target for the deposition of the thin films. Transport and magnetic measurements were performed on the polycrystal to compare the behaviour of the thin films with the bulk. Polycrystalline thin films of $Mn_3Sn$ were deposited on Pt coated $Al_2O_3$ substrates using Pulsed Laser Deposition (PLD). 5nm thin layer of Pt serves as a buffer layer for the growth and deposition stability (as the lattice mismatch between Pt and $Mn_3Sn$ in only around 0.12Å). Anomalous Hall effect was observed for the polycrystalline thin films with a large anomalous Hall conductivity of 65$(\Omega cm)^{-1}$ at 3K that remains greater than 60$(\Omega cm)^{-1}$ up to 20K. The Hall resistivity varied from 9n$\Omega$cm to 2.5n$\Omega$cm for temperatures of 3K to 150K. A negative transverse magnetoresistance (MR) was observed for all the measured temperatures in the range of 3K to 350K. Due to the presence of a very weak ferromagnetic moment arising from the geometrical frustration of the Mn spins, the magnetization measurements of the polycrystalline and thin film samples show a small spontaneous magnetization of less than 10emu/cc and a small coercivity that remains <0.3T at the lowest temperature and becomes vanishingly small at the higher temperatures.

**Results and discussion**

Fig. 1a shows the X-Ray diffraction pattern for the powdered polycrystalline $Mn_3Sn$ powder and $Mn_3Sn$ thin film on Pt coated $Al_2O_3$ substrate. The XRD pattern of the powder obtained from the polycrystal revealed the pure hexagonal single phase of $Mn_3Sn$ without any impurities. The Mn atoms form kagome planes in this material with Sn atoms at the centre of the hexagons as shown in fig.1c. The average thickness of the films was 110nm which was calculated by the Scanning Electron Microscopy (SEM) cross-sectional images as shown in Fig.1b.

The longitudinal resistivity and change in magnetization with applied field for the Mn$_3$Sn polycrystalline bulk sample are shown in Figure 2. Fig. 2a shows the longitudinal resistivity ($\rho_{xx}$) measured as a function of temperature in the range of 300K down to 3K. These measurements show a metallic behaviour with the maximum value of the resistivity, 167μΩcm at 300K and a value of 33μΩcm at 3K. Magnetic measurements on the polycrystal were performed in a Vibrating Sample Magnetometer (VSM). The change in magnetization with the applied magnetic field is shown in fig. 2b at different temperatures. Magnetization values (40emu/cc at 4T) at 10, 50 and 100K do not show any significant change, however, the value decreases (30emu/cc at 4T) at a higher temperature of 200K. The coercivity exhibited by the polycrystal is negligibly small (<0.05T), owing to a very weak ferromagnetic moment in this material. Inset of fig.2c shows the zoomed in version (up to the applied field range of 0.14T) of the magnetization loops.

For the thin films, the longitudinal resistivity ($\rho_{xx}$) as a function of temperature (RT) was measured from 400K to 3K as shown in the figure 3a. Inset shows the measurement configuration for the longitudinal resistivity measurements. Magnetic field was applied perpendicular to the direction of current for the MR measurements. The value of longitudinal resistivity for the thin films is significantly decreased as compared to the bulk polycrystalline sample ranging from 22μΩcm at 400K to 13μΩcm at 3K. The change in magnetoresistance (MR) with the applied field at different temperatures is shown in fig.3b. The value of the MR% remains negative at all the measured temperatures from 350K to 3K, with the decrease in the negative magnitude at higher temperatures. The MR was measured for the applied field of up to 5T, with the maximum negative value of 0.46% at 3K. The negative MR is observed in other topological magnetic materials and is attributed to the suppression of the spin scattering and fluctuations [21-23]. In our thin films, the negative MR at lower temperatures points to the fact that the fluctuations of Mn moments in the triangular arrangement are decreased. As the temperature increases and gets close to the Neel temperature, the MR starts to become positive as a result of more fluctuations and less ordering of the spin arrangement. This is also consistent with the large anomalous Hall conductivity measured for lower temperatures that decreases at higher temperatures.

Magnetic measurements on the thin film samples were carried out in the Vibrating Sample Magnetometer (VSM) mode of the PPMS to investigate the origin of the anomalous Hall effect. Fig. 3c shows the change in magnetization with the applied field at different temperatures. The diamagnetic contribution from the background was subtracted to get the magnetization loops for the thin films. The magnetization loops show small hysteresis, due to the weak ferromagnetic moment, like Hall measurements with the value of magnetization decreasing at the higher temperatures. The spontaneous magnetization values decrease from 10emu/cc to 2emu/cc as the temperature increases from 3K to 200K. The value of coercivity is very small and decreases from 0.2T at 3K to less than 0.1T at 200K, the hysteresis for the Hall measurements, however, becomes vanishingly small at 50K and higher temperatures. It is interesting to note here that the magnitude of the magnetic moment and coercivity remains almost same for both bulk and thin film samples.

The Hall effect measurements were carried out on thin film samples under the applied magnetic field of up to 3T at different temperatures as shown in figure 4. The inset of fig. 4a shows the measurement configuration for the Hall measurements carried out on the thin film samples. The applied magnetic field, the direction of current and the Hall voltage are perpendicular to each other. The Hall resistivity ($\rho_{xy}$) shows a significant jump around zero field with a small hysteresis. The value of $\rho_{xy}$ and the hysteresis decreases with the increase in temperature. The Hall measurements also indicate holes as being the charge carriers in the thin films.

The anomalous Hall resistivity $\rho^A_{xy}$ was calculated by the linear extrapolation of $\rho_{xy}$ at zero field. The variation of the anomalous Hall resistivity with temperature is shown in figure 5a. The magnitude of the Hall resistivity decreases from 9nΩcm to 2.5nΩcm with the increase in temperature from 3K to 150K. The magnitude of the anomalous Hall conductivity was calculated using the formula:

$$\sigma^A_{xy} = \frac{\rho^A_{xy}}{\left(\rho^{A2}_{xy} + \rho^2_{xx}\right)} \qquad (1)$$

Where $\rho^A_{xy}$ the anomalous Hall resistivity and $\rho_{xx}$ is the longitudinal resistivity.

The calculated values for Hall conductivity are large at lower temperatures ranging from 65(Ωcm)$^{-1}$ at 3K to 10(Ωcm)$^{-1}$ at 150K. It is worth noting that $\sigma^A_{xy} \geq 60$(Ωcm)$^{-1}$ up to 20K and starts to decrease at 50K. The values of anomalous Hall resistivity and conductivity decrease almost linearly with the temperature as shown in figs 5a and 5b.

Figure 5c and 5d show the variation of the anomalous Hall resistivity and conductivity with the square of their longitudinal counterparts $\rho_{xx}$ and $\sigma_{xx}$ that represents the quadratic dependence of $\sigma^A_{xy}$ and $\rho^A_{xy}$ on $\sigma_{xx}$ and $\rho_{xx}$. The resistivity plot shows a negative slope as there is an increase in the anomalous Hall resistivity at lower temperature when the longitudinal resistivity decreases. The skew scattering mechanism gives rise to a linear dependence of $\sigma^A_{xy}$ and $\rho^A_{xy}$ on $\sigma_{xx}$ and $\rho_{xx}$ in the AHE. The side jump mechanism and the intrinsic contribution from the Berry curvature both lead to the quadratic dependences. The AHE is put into three regimes with respect to the dependence of $\sigma^A_{xy}$ and the magnitude of the longitudinal conductivity. $\sigma_{xx} > 10^6$(Ωcm)$^{-1}$, $10^4$(Ωcm)$^{-1} < \sigma_{xx} < 10^6$(Ωcm)$^{-1}$ and $\sigma_{xx} < 10^6$ (Bad Metal regime) [24]. The effects from the skew scattering mechanism dominate in the first range, the second range in which the longitudinal conductivity for our thin film samples lies, is the regime in which the intrinsic effects are dominant.

As shown in fig. 5b, $\sigma^A_{xy}$ starts to decrease with the increase in temperature. The contribution from the impurity scatterings to the AHE would lead to its enhancement in the high temperature regime. The quadratic dependence of $\sigma^A_{xy}$ on $\sigma_{xx}$ and its decrease with the increase in temperature also point towards the intrinsic nature of the AHE observed in the thin films.

For the case of Mn$_3$Sn, the non-collinear antiferromagnetic order with the topological electronic band structure, the Berry curvature does not become zero in the momentum space and gives rise to a finite $\sigma^A_{xy}$. As discussed above, scattering does not seem to play a role in generating AHE in the thin films that points to its intrinsic origin.

**Conclusions**

Anomalous Hall effect was observed in polycrystalline thin films of Mn$_3$Sn (110nm) deposited on Pt coated Al$_2$O$_3$ substrates through PLD. A large anomalous Hall conductivity of 65 (Ωcm)$^{-1}$ was exhibited by the thin films at 3K, with the longitudinal conductivity lying in the regime where intrinsic effects are dominant. The anomalous Hall resistivity and conductivity decrease with the temperature dismissing the possibility of AHE due to extrinsic contributions that generally become relevant at higher temperatures. Polycrystalline bulk and thin film samples show the decrease of the magnetic moment and the coercivity with the increase in temperature. A negative transverse MR is observed for thin films at all measured temperatures up to 350K.

**Experimental Methods**

Polycrsytalline samples of $Mn_3Sn$ were prepared using the vertical furnace. 99.9% pure powders of Manganese (Mn) and Tin (Sn) were taken (with 10% excess Mn powder for the stability of the structure) and mixed and ground well in a mortar for ten minutes. Mixed powders were then sealed into a quartz tube and loaded into the furnace. The samples were then heated to 1000°C at the rate of 100°C/h and then kept at 1000°C for 6 hours, then cooled to 900°C at the rate of 1.25°C/h, which was then cooled down to room temperature under ambient conditions.

Round dark grey polycrystalline sample was obtained which was then cut into two pieces. One round tablet shaped piece was used as a target for the thin films. The other piece was further cut into small rectangular pieces and polished for the transport measurements. Small extra pieces were finely ground in a mortar to carry out the X ray diffraction measurements.

Thin films of $Mn_3Sn$ were deposited on $Al_2O_3$ substrate. A 5nm buffer layer of Pt was first deposited on the substrate to aid the growth of the film. Prepared polycrystal was used as a target for the deposition of the thin films. Films were deposited by using Pulsed Laser Deposition. The substrates were cut into rectangular pieces of ~4x6mm and ultrasonically cleaned in acetone. Films were deposited at a temperature of 400°C (for 40 minutes) and were kept at the same temperature post deposition (for 15 minutes) to ensure uniform and smooth surface of the thin films. The distance between the target and the substrate was 5cm. The laser power used for the deposition was $2 J/cm^2$. The base pressure before the annealing and deposition was set to $4 \times 10^{-4}$Pa.

**Acknowledgements**

This research is supported and funded by ARC Centre for Future Low Energy Electronics Technologies (ARC-FLEET), ARC Professional Future Fellowship (FT130100778).
We would like to thank Dr. Tony Romeo at Electron Microscopy Centre, AIIM, UOW, for his help in Scanning Electron Microscope (SEM) measurements

**Figures**

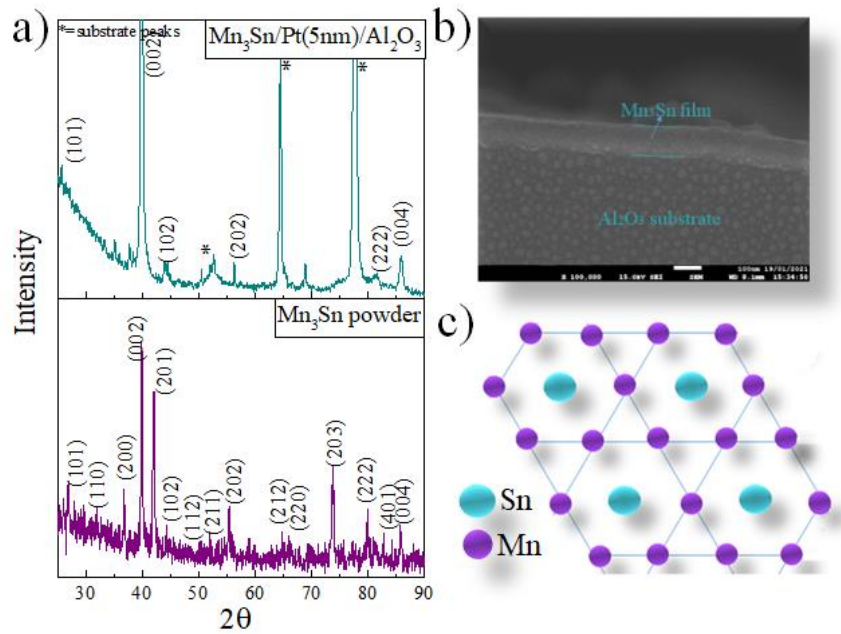

**Figure 1: a)** XRD images for: bottom) Mn₃Sn polycrystalline powder from the prepared target for thin film deposition. Top) Mn₃Sn thin film deposited on Al₂O₃ substrate with 5nm Pt coating. **b)** SEM cross-sectional scan of Mn₃Sn film on Pt coated substrate. **c)** Basal plane for Mn₃Sn crystal-Kagome lattice of Mn atoms at the corners of the hexagon and Sn atoms at the centre.

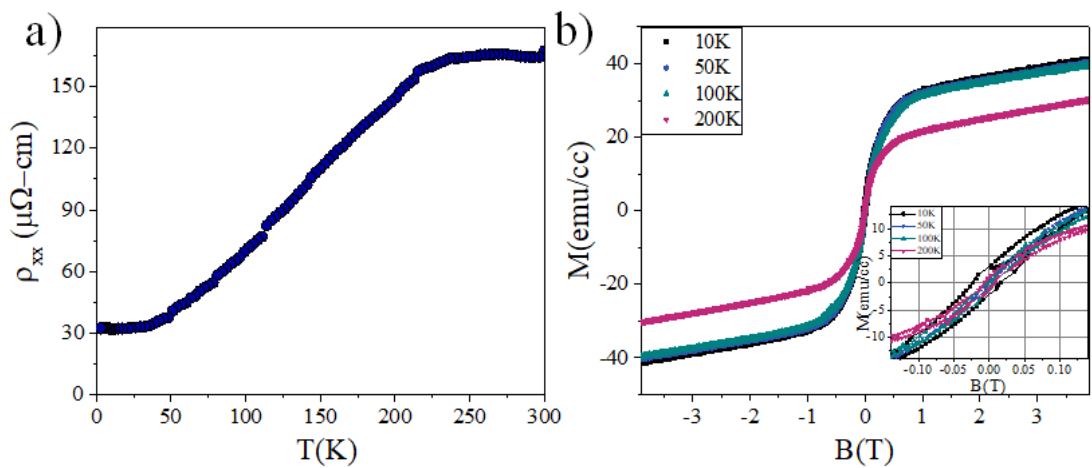

**Figure 2: a)** Variation in the longitudinal resistance with the change in temperature for Mn₃Sn polycrystal **b)** Variation in the magnetization with the applied field of up to 4T for Mn₃Sn polycrystal. Inset shows the zoomed in view of the magnetization loops. It can be seen that the coercivity remains <0.05T at 10K.

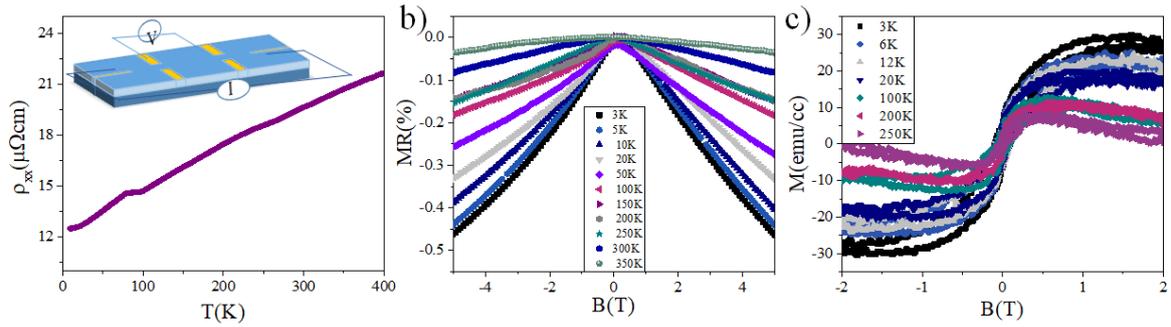

**Figure 3: a)** Variation of longitudinal resistivity of Mn$_3$Sn film with the temperature. Inset shows the schematics for the measurement configuration. **b)** Magnetoresistance (MR) of Mn$_3$Sn film at different temperatures from 3K to 350K. The MR (%) remains negative up to the highest measured temperature of 350K, however, the magnitude of MR decreases with the increase in temperature. **c)** The change in magnetization with the applied magnetic field of up to 2T. The magnitude of the magnetic moment and the coercivity decrease significantly at the higher temperatures.

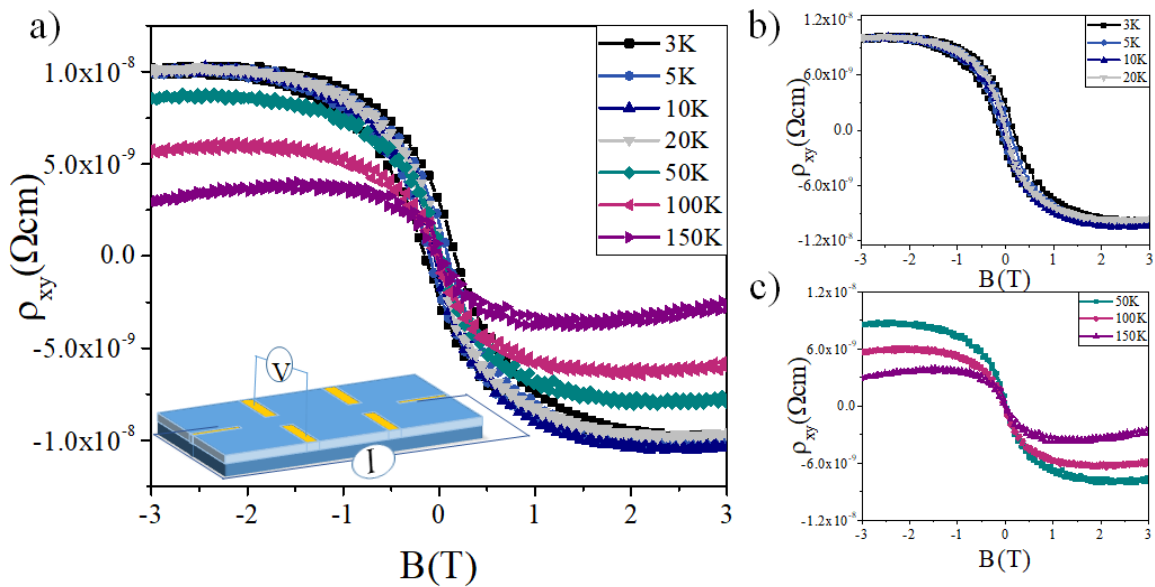

**Figure4: a)** The change in the Hall resistivity with the applied magnetic field at different temperatures. Inset shows the measurement configuration for the Hall measurements. The magnetic field was applied perpendicular to the plane of the sample. **b)** The change in the Hall resistivity with the applied field at lower temperatures 3,5,10 and 20K. The magnitude of the Hall resistivity doesn't vary much for this temperature range, however, the hysteresis is reduced as the temperature is increased. **c)** The change in the Hall resistivity with the applied field at higher temperatures of 50,100 and 150K. The magnitude of the Hall resistivity reduces considerably with negligible hysteresis at this temperature range.

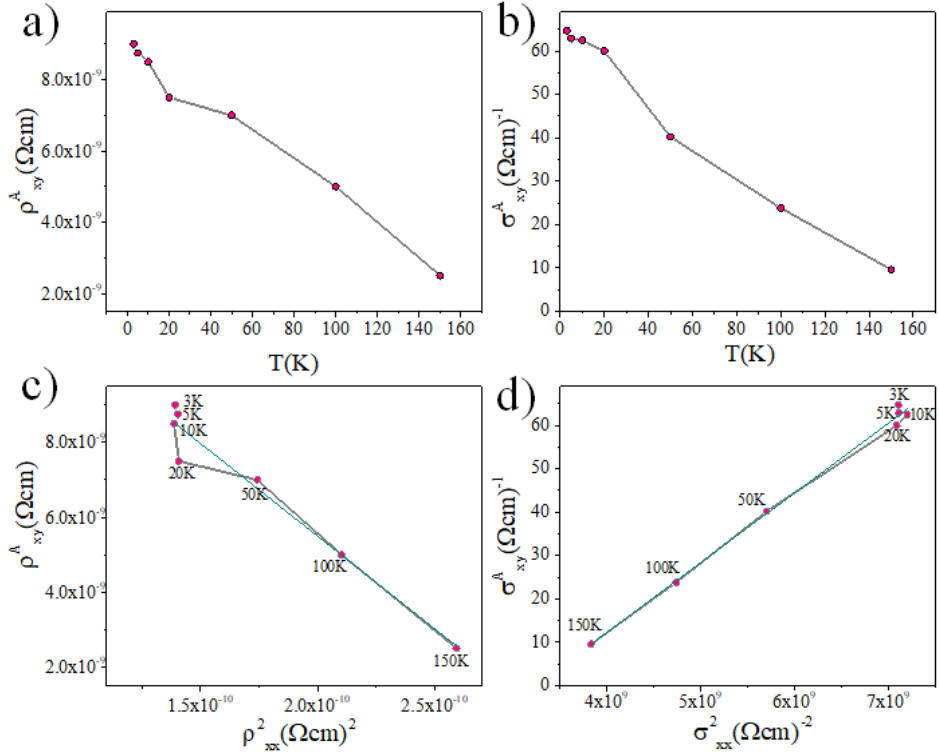

**Figure 5: a)** The change in anomalous Hall resistivity with temperature **b)** The change in anomalous Hall conductivity with temperature. Both the anomalous Hall conductivity and resistivity decrease almost linearly with the increase in temperature. **c)** The change in anomalous Hall resistivity with the square of longitudinal resistivity. **d)** The change in anomalous Hall conductivity with the square of longitudinal conductivity. (The green lines show the linear fits on the plots for **c** and **d**). $\rho^A_{xy}$ and $\sigma^A_{xy}$ show a quadratic dependence on $\rho_{xx}$ and $\sigma_{xx}$.